\documentclass[10pt,prl,nofootinbib,notitlepage,twocolumn,superscriptaddress]{revtex4-1}
\usepackage{amsmath}
\usepackage{amssymb,amsfonts}
\usepackage{graphicx}
\graphicspath{{pdfpics/}}
\usepackage{mathtools}
\usepackage{relsize}
\usepackage{lmodern}
\usepackage{slantsc}
\usepackage{scalefnt}
\usepackage{subfigure}
\usepackage{dsfont}
\usepackage{xspace}
\usepackage{boxedminipage}
\usepackage{ifpdf}
\usepackage{pgffor}
\usepackage{xifthen}
\usepackage{tensor}
\usepackage{array}
\usepackage{multirow}
\usepackage{tikz}
\usetikzlibrary{shapes.geometric}
\usepackage{color}
\usetikzlibrary{arrows,matrix,calc,scopes,decorations.markings}
\allowdisplaybreaks[1]
\usepackage{times}
\usepackage[colorlinks=true,linkcolor=blue,citecolor=blue,urlcolor=blue,pdfauthor={ },pdftitle={ },pdfsubject={ },pdfkeywords={ }]{hyperref}
\usepackage{epstopdf}
\usepackage{changes}

\usepackage{color}
\definecolor{dred}{rgb}{.8,0.2,.2}
\definecolor{ddred}{rgb}{.8,0.5,.5}
\definecolor{dblue}{rgb}{.2,0.2,.8}
\definecolor{dgreen}{rgb}{.2,0.5,.2}

\newcommand{\be}{\begin{equation}}
\newcommand{\ee}{\end{equation}}
\newcommand{\bse}{\begin{subequations}}
\newcommand{\ese}{\end{subequations}}
\newcommand{\ket}[1]{|{#1}\rangle}

\newcommand{\bra}[1]{\langle{#1}|}

\newcommand{\bpm}{\begin{pmatrix}}
\newcommand{\epm}{\end{pmatrix}}
\newcommand{\bmm}{\begin{matrix}}
\newcommand{\emm}{\end{matrix}}

\begin{document}

\title{Preserving Entanglement in a Solid-Spin System Using Quantum Autoencoders}

\author{Feifei Zhou}
\affiliation{Research Center for Quantum Sensing, Zhejiang Lab, Hangzhou, 311000, China}

\author{Yu Tian}
\affiliation{Shenzhen Institute for Quantum Science and Engineering and Department of Physics, Southern University of Science and Technology, Shenzhen 518055, China}

\author{Yumeng Song}
\affiliation{School of Physics, Hefei University of Technology, Hefei, Anhui 230009, China}

\author{Chudan Qiu}
\affiliation{Shenzhen Institute for Quantum Science and Engineering and Department of Physics, Southern University of Science and Technology, Shenzhen 518055, China}

\author{Xiangyu Wang}
\affiliation{Shenzhen Institute for Quantum Science and Engineering and Department of Physics, Southern University of Science and Technology, Shenzhen 518055, China}

\author{Mingti Zhou}
\affiliation{Research Center for Quantum Sensing, Zhejiang Lab, Hangzhou, 311000, China}

\author{Bing Chen}
\affiliation{School of Physics, Hefei University of Technology, Hefei, Anhui 230009, China}

\author{Nanyang Xu}
\email{nyxu@zhejianglab.edu.cn}
\affiliation{Research Center for Quantum Sensing, Zhejiang Lab, Hangzhou, 311000, China}

\author{Dawei Lu}
\email{ludw@sustech.edu.cn}
\affiliation{Shenzhen Institute for Quantum Science and Engineering and Department of Physics, Southern University of Science and Technology, Shenzhen 518055, China}

\begin{abstract}
Entanglement, as a key resource for modern quantum technologies, is extremely fragile due to the decoherence. Here, we show that a quantum autoencoder, which is trained to compress a particular set of quantum entangled states into a subspace that is robust to decoherence, can be employed to preserve entanglement. The training process is based on a hybrid quantum-classical approach to improve the efficiency in building the autoencoder and reduce the experimental errors during the optimization. Using nitrogen-vacancy centers in diamond, we demonstrate that the entangled states between the electron and nuclear spins can be encoded into the nucleus subspace which has much longer coherence time. As a result, lifetime of the Bell states in this solid-spin system is extended from $2.22\pm 0.43$ $\mu$s to $3.03\pm 0.56$ ms, yielding a three orders of magnitude improvement. The quantum autoencoder approach is universal, paving the way of utilizing long lifetime nuclear spins as immediate-access quantum memories in quantum information tasks.
\end{abstract}

\maketitle

\emph{Introduction.}---Entanglement lies at the heart of quantum information science. However, it is vulnerable due to the inevitable couplings between the quantum system and environment. To overcome decoherence, many strategies have been developed, including quantum error correction \cite{qec1, qec2, qec3, qec4, qec5}, dynamical decoupling \cite{dd1, dd2, dd3, dd4, dd5, dd6, wang2011preservation}, and decoherence-free subspaces (DFS) \cite{dfs1, dfs2, dfs3, dfs4}. In particular, the DFS approach encodes the entanglement information in a subspace where the decoherence effect is eliminated or minimized, leading to a much longer lifetime of experimental entanglement. Inspired by the DFS and emerging quantum machine learning techniques, we propose a new method to preserve entanglement by combining the quantum autoencoder \cite{quantum_autoencoder} and hybrid quantum-classical approach (HQCA) \cite{hqca1, hqca2, hqca3, hqca4, hqca5, xin2020quantum}.

The idea of autoencoders has been popular in the field of neural networks for decades \cite{deepnature, deepbook, AEscience}, typically for the purpose of dimensionality reduction \cite{AEscience}. It learns efficient data codings in an unsupervised manner, consisting of an encoder to learn a representation for a set of data and a decoder to generate from the reduced encoding a representation as close as possible to its original input. As an extension, quantum autoencoders have been recently proposed to compress a particular data set of quantum states into a lower-dimensional representation \cite{quantum_autoencoder}, and a proof-of-principle experiment that compresses the information of qutrit to qubit has been realized in optics \cite{qutrits}. Inspired by its demonstrated function in dimensionality reduction, we propose that a quantum autoencoder can be applied to preserve entanglement by compressing the high-dimensional entangled states into low-dimensional decoherence-invulnerable subspaces.

A classical autoencoder is often trained by artificial neural networks \cite{AEscience}. In the quantum realm, we propose to apply HQCA in the training process \cite{hqca1, hqca2, hqca3, hqca4, hqca5}. The underlying idea in HQCA is to exploit the particular advantages in quantum processors and classical computers to jointly perform gradient-based optimizations. Specifically, the cost function and gradient computations are accomplished on the quantum processor, while the remaining is done on a classical computer. Hence, HQCA offers a remarkable advantage in practice: as a ``bootstrap" quantum processor, it has the capability in compensating for unknown errors caused by its imperfect characterizations. This advantage has been verified in nuclear magnetic resonance \cite{bootstrapping} and electron paramagnetic resonance systems \cite{epr}. Moreover, in large-scale quantum systems, HQCA has ascendancy in computing efficiency compared to traditional methods \cite{hqca5,bootstrapping}. Therefore, incorporating HQCA into the quantum autoencoder training can in principle improve the efficiency and reduce experimental errors under experimental circumstances.

In this work, we demonstrate the scheme in a solid-spin system -- nitrogen-vacancy (NV) centers in diamond \cite{NVcenter,MLreadout, electricfield, tempnano, wavelet, vortexbeam,digitalphases, OTOC, ywnano, wigner}. In NV centers, the electron spin has fast control to realize reliable quantum gates and the nuclear spin has long coherence time as natual quantum memories \cite{reg_science, gate_pra, entanglement_science}. In addition, nuclear spins can be entangled with the electron to realize quantum-enhanced metrology \cite{chopped}. Therefore, the NV center is a good candidate to test the quantum autoencoder in the purpose of entanglement preservation. In experiment, we have trained a quantum autoencoder to extend the lifetime of Bell-type states by a three orders of magnitude in this solid-spin system. The results demonstrate the power of the HQCA-based quantum autoencoder in a mainstream quantum system for the purpose of entanglement preservation.

\emph{Quantum autoencoder.}---The classical autoencoder contains an encoder that maps the input into the code, and a decoder that maps the code to a reconstruction of the original input. For the purpose of dimensionality reduction, the code has a smaller dimension (referred to as the latent space) than the input. Both the encoder and decoder are usually trained by neural networks with single or multiple layers.

\begin{figure}[h]
\centering
\includegraphics[width=1\columnwidth]{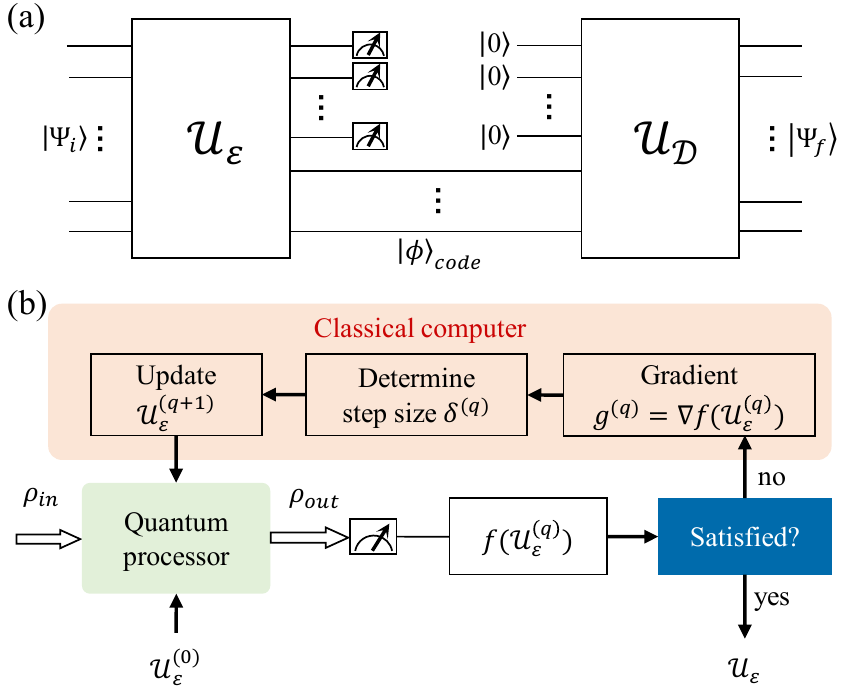}
\caption{ (a) Quantum autoencoder circuit. The information of $\ket{\Psi_{i}}$ can be encoded to the code state $\ket{\phi}_{code}$ via the encoder $\mathcal{U}_\mathcal{E}$. When needed, $\ket{\phi}_{code}$ can be reconstructed to $\ket{\Psi_{f}}$ by a decoder $\mathcal{U}_\mathcal{D}$. (b) Training process of the HQCA to optimize the encoder. $\rho_{in}$ is the input state of the encoder, and $\rho_{out}$ is the output state on the ancilla qubits. $f(\mathcal{U}_\mathcal{E}^{(q)})$ is the cost function, where $q$ is the current iterative number. If $f(\mathcal{U}_\mathcal{E}^{(q)})$ does not hit the preset value, the current trial $\mathcal{U}_\mathcal{E}^{(q)}$ will be updated by $\mathcal{U}_\mathcal{E}^{(q+1)}$ based on the measured gradients.}\label{fig1}
\end{figure}

For quantum autoencoders, the encoder and decoder are replaced by unitary transformations, as depicted in Fig. \ref{fig1}(a). The goal is to find an encoder, represented by some unitary $\mathcal{U}_\mathcal{E}$, that compresses the target information in $\ket{\Psi_{i}}$ into a smaller latent space. The other (ancilla) qubits out of the latent space can be traced over by post-selected measurements. To recover the information from the latent space, the ancilla are refreshed to a reference state $\ket{00...0}$ and a decoder $\mathcal{U}_\mathcal{D}$ is applied to maximize the similarity between the output $\ket{\Psi_{f}}$ and the original input $\ket{\Psi_{i}}$. The structure of the encoder and decoder can be programmable quantum circuits \cite{quantum_autoencoder}.

This model can be used to preserve entanglement. Suppose $\ket{\Psi_{i}}$ is an entangled state, and there exists a latent subspace where the information is more invulnerable to decoherence, e.g., in the DFS or simply in some qubits that have longer coherence time. The aim is to find an encoder $\mathcal{U}_\mathcal{E}$, such that $\ket{\Psi_{i}}$ will be evolved to $\ket{00...0}_{anc}\otimes \ket{\phi}_{code}$ after encoding, where $\ket{00...0}_{anc}$ is the state of the ancilla qubits and $\ket{\phi}_{code}$ is the code state in the latent space. As the latent space is robust to decoherence errors, it is a reliable quantum memory to store the quantum information of $\ket{\phi}_{code}$. To recover the original entangled states, one reinitializes the ancilla qubits to $\ket{00...0}$ and apply the decoder $\mathcal{U}_\mathcal{D} = \mathcal{U}_\mathcal{E} ^ {\dagger}$. In the ideal case, i.e., if no errors are involved during the operations, the ideal output would reconstruct the input perfectly.

The crucial part is how to train the encoder $\mathcal{U}_\mathcal{E}$. As the encoder is some unitary, its structure can be defined by a parameterized quantum circuit (PQC), so that the number of parameters and gates scale polynomially with the number of input qubits \cite{quantum_autoencoder,xin2021experimental}. This PQC encoder can be trained by maximizing the cost function $f(\mathcal{U}_\mathcal{E}^{(q)})$ via iteratively updating $\mathcal{U}_\mathcal{E}$; see Fig. \ref{fig1}(b). This optimization is implemented using HQCA \cite{supplemental_material}. As mentioned before, only the most time-consuming parts, such as the measurement of the cost function and gradients, are accomplished on the quantum processor. This drastically reduces the optimization complexity in large-scale systems. In addition, as the outcomes are measured on the quantum processor, the optimization process can automatically account for certain errors during the evolution. These two advantages of HQCA make it a perfect candidate for the encoder optimization problem.

\begin{figure}[ht]
	\centering
	\includegraphics[width=1\columnwidth]{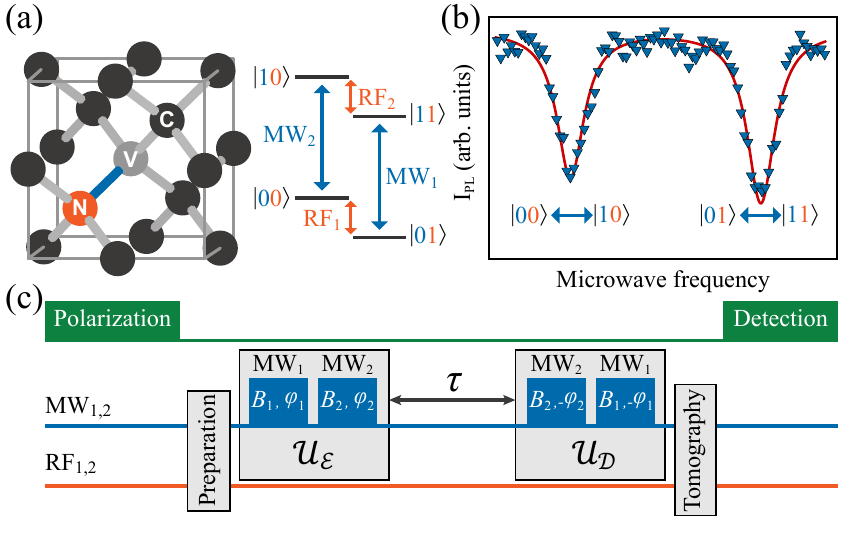}
	\caption{(a) Schematics of an NV center in diamond and partial energy-level with a substitutional $^{14}$N nuclear spin. Transitions between electron (nuclear) energy levels are excited via MW$_1$ (RF$_1$) and MW$_2$ (RF$_2$), respectively. (b) ODMR spectrum of the electron spin in experiment, where frequencies of MW$_1$ and MW$_2$ are determined according to two resonance peaks. (c) Pulse sequence to implement the quantum autoencoder experiment. The encoder $\mathcal{U}_\mathcal{E}$ and decoder $\mathcal{U}_\mathcal{D}$ only involve microwave pulses on the electron spin. $\tau$ is a tunable free evolution time.}
	\label{fig2}
\end{figure}

\emph{Experiment.}---We use the NV center in diamond to demonstrate the quantum autoencoder scheme in preserving entanglement. The NV center involves one electron spin with fast operations and neighbouring nuclear spins with extremely long coherence time, forming a hybrid solid-spin system as shown in Fig. \ref{fig2}(a). A widely recognized advantage of the NV center is that the electron spin can play as an actuator qubit to indirectly control the nuclei, while the nuclei are employed as quantum memories \cite{memory1, memory2, memory3, memory4}. This feature is a perfect match to our scheme, as entangled states between the electron and nuclear spins can be encoded into the nuclear subspace without operating the nuclear spins. Furthermore, this complex solid-spin system with hard-to-characterize noises offers a testbed to demonstrate the advantage of HQCA optimization in reducing experimental errors.

Here, we consider the Bell-type states
\be
\ket{\Psi_{i}} = \alpha \ket{00} + \beta \left | 11 \right \rangle(\alpha, \beta \in \mathbb{C}; |\alpha|^2 + |\beta|^2 = 1)
\label{bell}
\ee
between the electron and the $^{14}$N nucleus as the target entangled states. As the most important entangled states in quantum information, Bell states are representative to demonstrate this scheme. The scheme is general, which is capable of tackling more complex and higher-dimensional entangled states; see the Supplemental Information  \cite{supplemental_material}.

Experiments are performed with the NV center in a bulk diamond on a home-built optically detected magnetic resonance (ODMR) system \cite{pulse_width_induced}. A green laser of 532 nm modulated by acousto-optic modulator is applied via the confocal microscopy system and the fluorescence ranging from 650 to 800 nm is detected using the avalanche photodiode. A static magnetic field $B_z \approx 52$ mT is applied along the NV axis to split the $\ket{m_S = \pm 1}$ and enable the optical polarization of both spins via the excited state level anti-crossing \cite{eslac}. The microwave signals with different frequencies and phases are generated from an IQ-modulation system, consisting of an arbitrary waveform generator (Tektronix AWG610) and a vector signal generator (Rohde$\&$Schwarz SMIQ03B). The AC magnetic field is radiated to the sample via a slot-line structure with an $\Omega$-type ring on the coverslip with the thickness of 100 $\mu$m.

The energy levels are shown in Fig. \ref{fig2}(a). We denote $\textbf{S}$ and $\textbf{I}$ as the electron and nuclear spins, respectively. The Hamiltonian can be written as
\begin{equation}
\mathcal{H} = D S_z^2 + g_e \mu_e \textbf{B} \cdot \textbf{S} + \textbf{S} \cdot \textbf{A} \cdot \textbf{I} + Q I_z^2 + g_n\mu_n \textbf{B} \cdot \textbf{I},
\end{equation}
where $D \approx 2.87$ GHz is the zero-field splitting, $\textbf{A}$ is the tensor of hyperfine interaction, $Q$ is the nuclear quadruple interaction, $\mu_e$ and $\mu_n$ are corresponding magnetons, and $g_e$ and $g_n$ are the $g$-factors, respectively.  When pumped with a 532 nm laser pulse at a 52 mT static magnetic field, excited state level anti-crossing happens, leading to simultaneous polarization of the electron and nuclear spins \cite{eslac}. After that, we calibrate the resonant frequencies for both spins using the electron-nuclear double resonance (ENDOR) and ODMR spectra, respectively. The ODMR spectrum is shown in Fig. \ref{fig2}(b), and $\pi$ pulses for RF$_1$, RF$_2$, MW$_1$, and MW$_2$ are calibrated via Rabi experiments \cite{supplemental_material}.

\emph{Results.}---To find the encoder for the Bell-type states in Eq. (\ref{bell}), we separate the entire optimization into two processes for different initial states $\ket{00}$ and $\ket{11}$, respectively. According to the superposition principle, if the encoder can compress both $\ket{00}$ and $\ket{11}$, it works for any superposition of them. The encoder, represented by the PQC in Fig. \ref{fig2}(c), is chosen as a sequence of two microwave pulses: MW$_1$ with amplitude $B_1$ and phase $\phi_1$, and MW$_2$ with $B_2$ and $\phi_2$. Namely, pulses are only applied on the electron spin to guarantee fast operations. The duration of each microwave pulse is fixed as $t=800$ ns. The unitary form of the encoder is thus $\mathcal{U}_\mathcal{E}=U_{\text{MW}_{\text{2}}}U_{\text{MW}_{\text{1}}}$, where $U_{\text{MW}_{\text{1}}}=\text{exp}[{-i \pi t B_1 (\text{cos} \phi_1 \sigma_x + \text{sin} \phi_1 \sigma_y)  \otimes \ket{1}\bra{1}}]$ and $U_{\text{MW}_{\text{2}}}=\text{exp}[{-i \pi t B_2 (\text{cos} \phi_2 \sigma_x + \text{sin} \phi_2 \sigma_y)  \otimes \ket{0}\bra{0}}]$.

The HQCA optimization of the encoder goes as follows. The values of amplitudes $B_1$ and $B_2$ are represented by the voltage peak-peak of the AWG. We set $B_1 = 0.02$ V, $B_2 = 0.1$ V, and $\phi_1=\phi_2= \pi/4$ as the initial guess, and then apply this trial $\mathcal{U}_\mathcal{E}$ to the input states $\ket{00}$ and $\ket{11}$, respectively. The cost function $f(\mathcal{U}_\mathcal{E})$ is defined by the average probability $P$ that the electron spin occupies $\ket{0}$ in the two experiments, which can be obtained by measuring the population of the electron spin. When $P$ approximates 1, it means the encoder $\mathcal{U}_\mathcal{E}$ can evolve both $\ket{00}$ and $\ket{11}$ to a \emph{code} in which the electron is $\ket{0}$. Therefore, for any linear combination of $\ket{00}$ and $\ket{11}$, the encoder transforms it to a certain state where the electron is $\ket{0}$. This is the aim of the quantum autoencoder, because the electron spin can be discarded and reinitialized to $\ket{0}$ when needed.

The encoder is optimized iteratively by measuring the amplitude's gradient. We set a tunable step size $\delta^{(q)}$, which starts from 0.05 V and varies according to the current variation of the cost function $f(\mathcal{U}_\mathcal{E}^{(q)})$ \cite{supplemental_material}. $q$ is the current iterative number. For example, to update the amplitude $B_1^{(q)}$ of MW$_1$, we change its value by $\delta^{(q)}$ and measure the new cost function, while all other parameters are fixed. The gradient is thus computed by $g^{(q)}=\nabla f(\mathcal{U}_\mathcal{E}^{(q)})$. The amplitude is updated by $B_1^{(q+1)} = B_1^{(q)} + \epsilon g^{(q)}$, where $\epsilon = 0.006$ is the learning rate. This gradient measurement as well as the update is repeated for all parameters before the next iteration.

The measurement of the cost function $f(\mathcal{U}_\mathcal{E}^{(q)})$ and gradient $g^{(q)}$ are both implemented on the NV center. The advantage in reducing experimental errors is remarkable in this experiment, because there are many error sources that need prior characterizations in this solid-spin system, such as the control imperfections induced by the nonlinearity of devices or the decoherence caused by the nuclear spin bath. The HQCA avoids preemptive corrections of these errors while maintaining high optimization fidelity, as shown in the following.

\begin{figure}[ht]
	\centering
	\includegraphics[width=1\columnwidth]{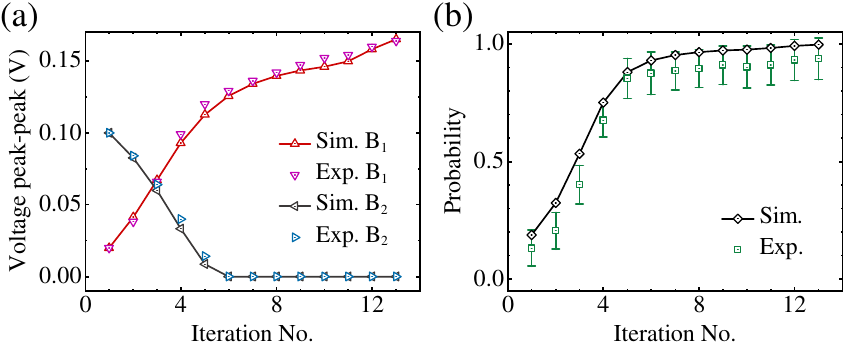}
	\caption{(a) Variations of $B_1$ and $B_2$ during the optimization. The values of $B_1$ and $B_2$ are represented by the voltage peak-peak of the AWG. Experimental results are in good agreement with the simulations. (b) Average probability that the electron spin occupies $\ket{0}$ after the encoder during the training. The probability is measured via state tomography on diagonal elements. The discrepancy between the experiment and simulation is mainly due to the decoherence effect in the readout stage.}
	\label{fig3}
\end{figure}

The quantum autoencoder is successfully trained after 13 iterations in experiment. The final values of  $B_1$ and $B_2$ are 0.164 V and 0 V, which match well with the simulations, as shown in Fig. \ref{fig3}(a). The cost function $f(\mathcal{U}_\mathcal{E})$, i.e., probability of the electron spin occupying $\ket{0}$, converges to over 0.93; see Fig. \ref{fig3}(b).  Error propagation method \cite{wigner} is used to draw the error bars. The phases $\phi_1$ and $\phi_2$ cast no impact on the cost function, so their values remain unchanged during the training.

\begin{figure}[ht]
	\centering
	\includegraphics[width=1\columnwidth]{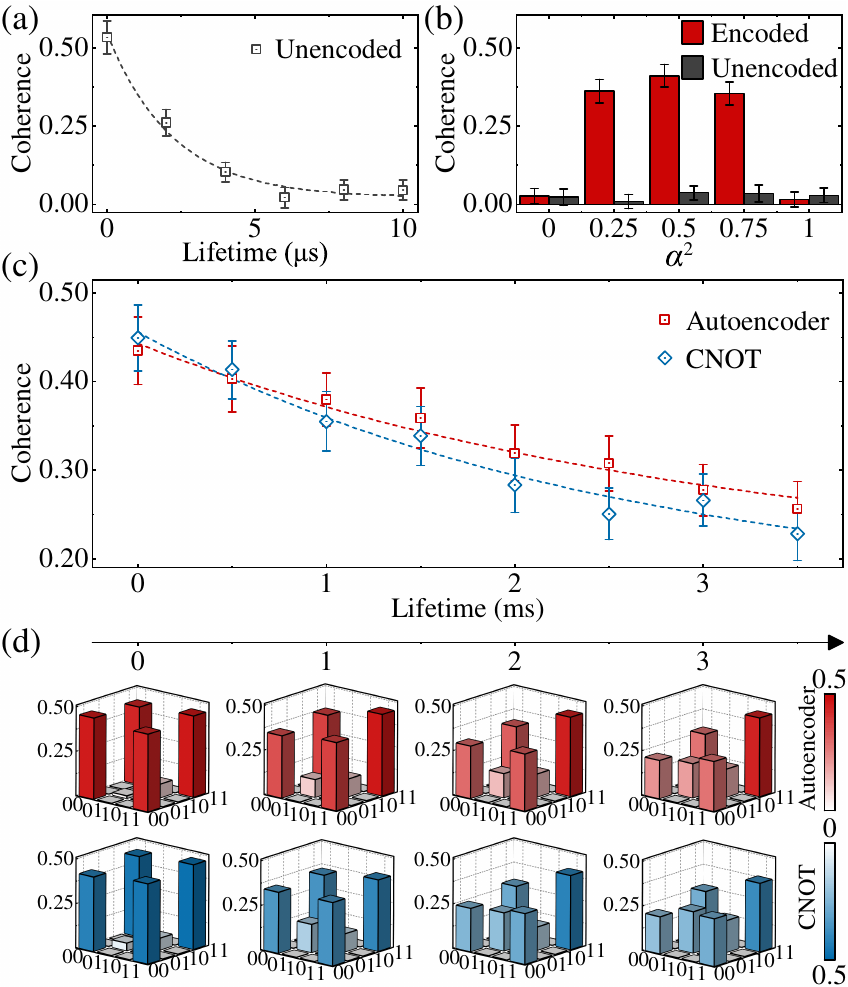}
	\caption{(a) Lifetime of the Bell state without encoding. The lifetime of uncompressed Bell state is $2.22(43)$ $\mu$s. (b) For other entangled states with different coefficients $\alpha^2 =  \{0, 0.25, 0.5, 0.75, 1\}$, at $\tau = 300$ $\mu$s, the residual entanglement with encoding is much larger than the case without encoding. (c) Lifetime of the Bell state compressed by the CNOT gate and the autoencoder. The estimated lifetime for the two cases are about $2.53(76)$ ms and $3.03(56)$ ms, respectively. The density matrices at integer milliseconds are also plotted in (d) for reference. }
	\label{fig4}
\end{figure}

To verify the effect of the trained autoencoder, we conduct three groups of experiments and plot the amount of double-quantum coherence $\ket{00}\bra{11}$ with the growth of time. First, we let the Bell state $(\ket{00}+\ket{11})/\sqrt{2}$ directly undergo free evolution, and measure its fidelity decay for every 2 $\mu$s. Not surprisingly, the entanglement dies quickly in a few microseconds due to its fragility; see Fig. \ref{fig4}(a). The fitted entanglement lifetime is about $2.22(43)$ $\mu$s \cite{supplemental_material}. Second, we apply a traditional CNOT gate to disentangle the Bell state before the free evolution. After a certain time, we apply another CNOT gate to recover the entangled state and measure the fidelity decay. The lifetime is extended to $2.53(76)$ ms. The improvement is significant, as a CNOT gate transforms the Bell state to $\ket{00}$ which has much better coherence time. Third, we apply the encoder before the free evolution, allowing the entanglement to be compressed into the nuclear subspace. After applying the decoder, the entanglement lifetime is measured to be $3.03(56)$ ms. This is the best performance among all three cases, which is three orders of magnitude larger than the unencoding case and 19.8\% improvement compared to the second case, respectively.

Moreover, we test the quantum autoencoder for a particular set of entangled states. From our training process, we know that the encoder has the capability to compress all Bell-type states spanned by $\ket{00}$ and $\ket{11}$. In experiment, we vary the state form by choosing different coefficients $\alpha^2 =  \{0, 0.25, 0.5, 0.75, 1\}$, and test the autoencoder performance for each coefficient. We measure the residual $\ket{00}\bra{11}$ after $\tau = 300$ $\mu$s evolution. From Fig. \ref{fig4}(b), we see that the residual $\ket{00}\bra{11}$ is much larger for all entangled states with encoding, indicating the generality of the trained autoencoder.

\emph{Discussion.}---The reason that the quantum autoencoder outperforms the CNOT-decoupling method should be attributed to the HQCA approach, where errors are incorporated and corrected automatically during the optimization. Traditional CNOT gates suffer control imprecision problem, where the dominating error is the non-linearity of electronic devices. Although this type of error can be partially corrected by measuring the amplitude and voltage peak-peak relation, the HQCA-based autoencoder approach requires no prior knowledge about the noise model, which avoids its characterization in advance.

The HQCA-based quantum autoencoder is general. It can in principle compress any set of states into a latent subspace as long as the \emph{degrees of freedom} before and after compression are matching \cite{quantum_autoencoder}. We have performed numerical simulations for more complex entangled states including the Greenberger-Horne-Zeilinger state, W-state, and 4-qubit cat-state \cite{supplemental_material}. The results demonstrate the solidity of the quantum autoencoder optimization. In this work, there are two reasons that we intentionally choose the input as Bell-type entangled states: they are the most important entangled states in quantum information, and they can be easily disentangled by a CNOT gate. The latter reason is important as it enables a fair comparison in performance with the autoencoder approach, in particular to exhibit the advantage of HQCA.

\emph{Conclusion.}---Recent developments in machine learning provide many powerful tools for tackling quantum tasks. In this work, we focus on a central issue in quantum information processing, i.e., to preserve fragile entangled states in practice. We adopt the quantum autoencoder idea and design a HQCA method to optimize the encoder. The codes of the target entangled states are stored in a subspace with much longer coherence time, thus enabling the extension of the entanglement lifetime. We conduct experiments in the hybrid solid-spin NV center system, and implement the compression for Bell-type entangled states. The entanglement lifetime is prolonged by three orders of magnitude compared to the unencoding case, and by 19.8\% compared to the case of CNOT decoupling. The HQCA-based quantum autoencoder is general, and we anticipate it to be a more useful tool in other quantum information tasks, such as quantum data compressor or denoiser \cite{denoiser}.

\begin{acknowledgments}
This work is supported by the National Key Research and Development Program of China (Grants No. 2018YFA0306600, 2019YFA0308100, and 2020YFA0309400), the National Natural Science Foundation of China (12075110, 11975117, 11905099, 11875159  and U1801661),  the Guangdong Basic and Applied Basic Research Foundation (2019A1515011383), the Guangdong International Collaboration Program (2020A0505100001), the Science, Technology and Innovation Commission of Shenzhen Municipality (ZDSYS20170303165926217, KQTD20190929173815000, JCYJ20200109140803865, JCYJ20170412152620376 and JCYJ20180302174036418),  the Pengcheng Scholars, the Guangdong Innovative and Entrepreneurial Research Team Program (2019ZT08C044), the Guangdong Provincial Key Laboratory (2019B121203002), the Major Scientific Project of Zhejiang Laboratory (2019MB0AE03), and the Fundamental Research Funds for the Central Universities.

F. Z. and Y. T. contributed equally to this work.
\end{acknowledgments}

\clearpage
\begin{center}
\widetext{\textbf{\large Preserving Entanglement in a Solid-Spin System Using Quantum Autoencoders: 
\\Supplemental Information}}\\
~\\
Feifei Zhou,$^1$ Yu Tian,$^2$ Yumeng Song,$^3$ Chudan Qiu,$^2$ Xiangyu Wang,$^2$ \\
Mingti Zhou,$^1$ Bing Chen,$^3$ Nanyang Xu,$^{1,*}$ and Dawei Lu$^{2,\dagger}$\\
~\\
$^1$\textit{Research Center for Quantum Sensing, Zhejiang Lab, Hangzhou, 311000, China}\\
$^2$\textit{Shenzhen Institute for Quantum Science and Engineering and Department of Physics, \\Southern University of Science and Technology, Shenzhen 518055, China}\\
$^3$\textit{School of Physics, Hefei University of Technology, Hefei, Anhui 230009, China}\\
~\\
*~E-mail Address: nyxu@zhejianglab.edu.cn\\
$\dagger$~E-mail Address: ludw@sustech.edu.cn\\
\end{center}
\clearpage

\section{optimization of the quantum autoencoder}
Quantum autoencoder is a novel machine learning method applied in the quantum realm, and optimization of the autoencoder is based on the hybrid quantum-classical approach (HQCA). HQCA uses the quantum processor to optimize itself. Thus it can incorporate some unknown errors in this complex solid-spin system during optimization, resulting in a better result compared to the classical optimization approach. Our scheme is to train the quantum autoencoder using HQCA, and use it to search parameters for compression of two-qubit entanglement.

In the compression process of entangled state between the electron and nuclear spins in nitrogen-vacancy (NV) center, we label the electron spin as qubit 1, and the nuclear spin $^{14}$N as qubit 2. The key point is to find some unitary $\mathcal{U}$ in terms of parametrized control pulses, such that all entangled states of the form $\alpha \left | 00 \right \rangle + \beta \left | 11 \right \rangle(\alpha, \beta \in \mathbb{C}; |\alpha|^2 + |\beta|^2 = 1)$ can be compressed into a product state $\left | 0 \right \rangle \otimes \left | \psi \right \rangle$. $\left | \psi \right \rangle$ is arbitrary since the information is already encoded into the nuclear spin, meaning that its lifetime is mainly limited by the relaxation time of the nuclear spin. The electron spin state $\left | 0 \right \rangle$ can be discarded after the autoencoder. To recover the original entangled state, we can reinitialize the electron spin to $\left | 0 \right \rangle$ and apply $\mathcal{U}^{\dagger}$ on $\left | 0 \right \rangle \otimes \left | \psi \right \rangle$.

Now we introduce how a standard quantum autoencoder works in the compression of quantum states. In the iteration $q$, we prepare the entangled state $\rho_{in}$ and determine the current encoder $\mathcal{U}^{(q)}$. When $q=1$, the initial encoder can be generated by a random guess. The result $\rho_{out}$ is calculated on a classical computer, and the value of the cost function $f(\mathcal{U}^{(q)})$ is extracted. If the value does not hit the preset value, go to the next iteration and update the encoder along the current gradient direction.

In our scheme of preserving entanglement using quantum autoencoder, we combine the HQCA to train the encoder. Here, we use the NV center experiment to describe how the scheme is implemented. As discussed in the main text, we adopt two microwave (MW) pulses MW$_1$ and MW$_2$ to parametrize the encoder. The duration of each pulse is fixed by 800 ns, and the amplitudes and phases are set as tunable parameters. The value of amplitudes $B_{\text{1}}$ and $B_{\text{2}}$ are represented by the voltage peak-peak of arbitrary waveform generator (AWG). Initial values of $B_\text{1}$, $B_\text{2}$, $\phi_\text{1}$, and $\phi_\text{2}$ are respectively fixed to be 0.02 V, 0.1 V, $\pi/\text{4}$ and $\pi/\text{4}$ as the initial guess. The optimization goes as follows:
\begin{itemize}
\item[1.] Prepare $\ket{00}$. Apply the two MW pulses with current values of $B_\text{1}$, $B_\text{2}$, $\phi_\text{1}$, and $\phi_\text{2}$. Measure the probability that the electron spin occupies $\left | 0 \right \rangle$ via the tomography of diagonal elements. Call it $P_{\ket{00}}$. Then prepare $\ket{11}$ and measure $P_{\ket{11}}$. The average probability is thus $P_{\left | 00 \right \rangle + \left | 01 \right \rangle}$.
\item[2.] Increase $B_\text{1}$ by the step $\Delta B = 0.05$ V and keep $B_\text{2}$ unchanged. Repeat Step 1 and measure the probability. Calculate how much it increases/decreases. Call it $\Delta P_{\left | 00 \right \rangle + \left | 01 \right \rangle}$.
\item[3.] The changes of probability by changing $B_\text{1}$ is $\Delta P_{\left | 00 \right \rangle + \left | 01 \right \rangle}$. If $|\Delta P_{\left | 00 \right \rangle + \left | 01 \right \rangle}$ is smaller than 2 $\%$, set the new $\Delta B$ to be half of its current value for the next iteration.
\item[4.] So the gradient is $\Delta G = \Delta P_{\left | 00 \right \rangle + \left | 01 \right \rangle} / \Delta B$. For the next iteration, update $B_1 \rightarrow B_1 + \epsilon \Delta G$. Considering the real situation, if the calculation result of the next $B_\text{1}$ is negative, set $B_\text{1}$ as zero. Here we fix the learning rate $\epsilon$ to be 0.006.
\item[5.] Reset all parameters to the initial values and repeat Step 1 to 4 to find how to update $B_\text{2}$. Since the probability does not rely on the phases, we do not consider the optimizations of $\phi_\text{1}$ and $\phi_\text{2}$.
\item[6.] One iteration is done. Repeat several iterations until the final probability $P_{\left | 00 \right \rangle + \left | 01 \right \rangle}$ that the electron spin occupies $\left | 0 \right \rangle$ approaches 1 and gets stable.
\end{itemize}

After 13 iterations, the probability of $\left | 0 \right \rangle _\text{e}$ is beyond 0.93 and gets stable. The quantum autoencoder is thus optimized.

\section{preservation of the entanglement}
After optimizing the parameters of the autoencoder, we conduct three groups of experiments to verify its performance and make comparisons with reference experiments, as shown in Fig. \ref{figs2}(d).
\begin{itemize}
	\item[1.] Prepare the Bell state $\frac{1}{\sqrt{2}} (\left | 00 \right \rangle + \left | 11 \right \rangle)$.
	\item[2.] Exp \emph{i}. Do nothing on the state and measure its fidelity decay for every 2 $\mu$s.
	\item[3.] Exp \emph{ii}. Use a controlled-NOT (CNOT) gate to decouple the state first, do nothing and let it undergo free evolution. For every 500 $\mu$s, apply the CNOT again to recover the entangled state and measure its instantaneous fidelity.
	\item[4.] Exp \emph{iii}. Use the quantum autoencoder to encode the state first, do nothing and let it undergo free evolution. For every 500 $\mu$s, apply the decoder to recover the state and measure its instantaneous fidelity.
\end{itemize}

The experimental results are fitted using the exponential decay function $y = y_0 + A_0 e^{-x / t}$ to extract the lifetime of entanglement. The lifetime of entanglement for all three groups of experiments are 2.22 $\pm$ 0.43 $\mu$s, 2.53 $\pm$ 0.76 ms, and 3.0352 $\pm$ 0.56 ms, respectively. The error bars are calculated using the error propagation method. All the experiments are repeated by 3 $\times$ $\text{10}^6$ times.

\section{experimental setup}
In this work, we perform experiments with an NV center in a bulk diamond on a home-bulit optically detected magnetic resonance (ODMR) system. A green laser of 532 nm modulated by acousto-optic modulator (AOM) is applied to the sample via the confocal microscopy system and the fluorescence ranging from 650 to 800 nm is detected by using the avalanche photodiode (APD). The laser is focused by a 60$\times$ oil-immersed objective with NA of 1.42. A static magnetic field $B_z \approx 52$ mT provided by a columnar neodymium magnet is applied along the NV axis to split the $m_S = \pm 1$ sublevels and enable the excited state level anti-crossing (esLAC). The MW and radio-frequency (RF) signals with different frequencies and phases are generated from an IQ-modulation system, mainly consist of an AWG (Tektronix AWG610) and a vector signal generator (Rohde$\&$Schwarz SMIQ03B). The AC magnetic field is radiated to the sample via a slot-line MW delivery with an $\Omega$-type ring on the coverslip with the thickness of 100 $\mu$m, as shown in Fig.~\ref{figs1}(a).

\renewcommand{\thefigure}{S1}
\begin{figure}[ht]
	\centering
	\includegraphics[width=0.8\textwidth]{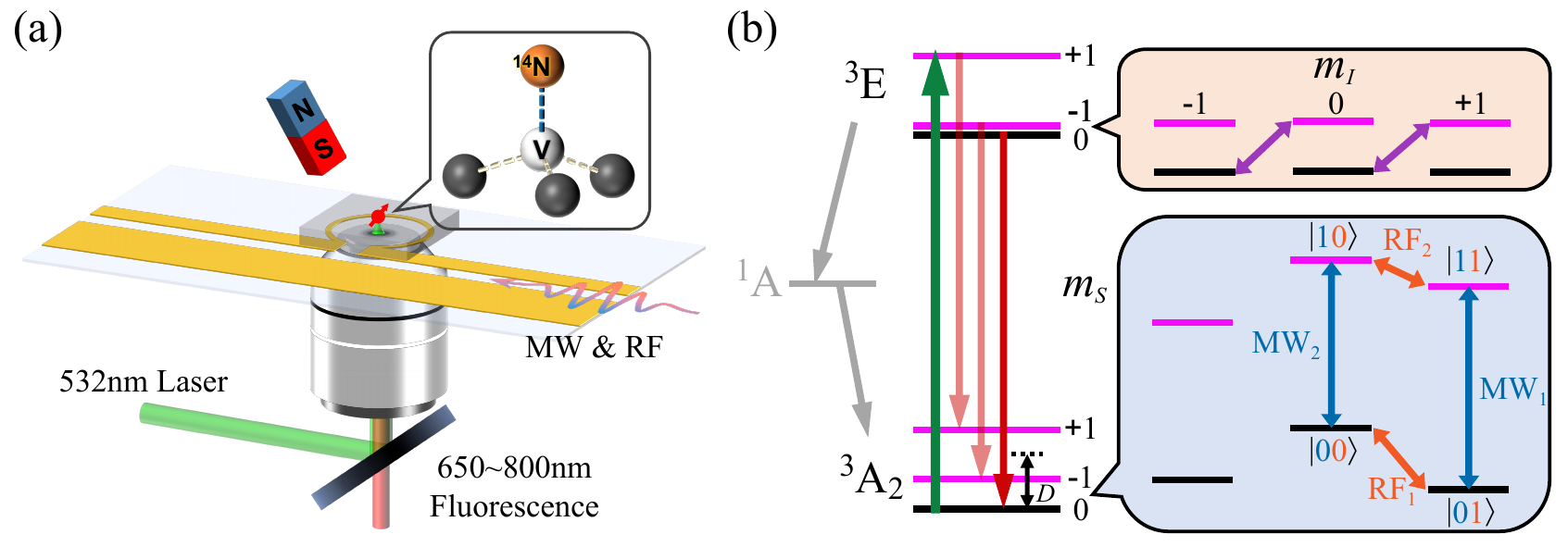}
	\caption{(a) shows the partial appratus of confocal microscopy system. (b) shows the energy level structure of electron spin in NV center with the hyperfine interaction of $^{14}$N nucleus. Here the gray arrow represents the intersystem crossing, and the purple arrows represent the spin flip-flops of $\left | 0, -1 \right \rangle \leftrightarrow \left | -1, 0 \right \rangle$ and $\left | 0, 0 \right \rangle \leftrightarrow \left | -1, +1 \right \rangle$. MW$_{1, 2}$ and RF$_{1, 2}$ correspond to the transitions of $\left | 0 \right \rangle _{\text{e}} \leftrightarrow \left | 1 \right \rangle _{\text{e}}$ and $\left | 0 \right \rangle _{\text{n}} \leftrightarrow \left | 1 \right \rangle _{\text{n}}$, respectively.}
	\label{figs1}
\end{figure}

The energy level of electron spin in NV center coupling with a $^{14}$N nucleus is shown in Fig. \ref{figs1}(b). A 532 nm laser pulse (green arrow) enables the transition between the ground spin-triplet ($^3$A$_2$) states and the excited states ($^3$E), while the excited states release fluorescence (red arrows) when it radiatively relaxes down to the ground states. Meanwhile, the electron spin on the excited $m_S = \pm 1$ state can pass through intersystem crossing (ISC) with a small probability and decay to the ground $m_S = 0$ state, which allows optical polarization of electron spin with 532 nm laser.

Due to coupling to $^{14}$N nucleus, the spin-Hamiltonian of the NV center can be written as the sum of zero-field splitting term, electron spin and nuclear spin Zeeman splitting terms, the hyperfine interaction term (hyperfine splitting tensor $\textbf{A}$), and nuclear quadrupole interaction ($Q$):
\renewcommand{\theequation}{S1}
\begin{equation}
\mathcal{H} = D S_z^2 + g_e \mu_e \textbf{B} \cdot \textbf{S} + \textbf{S} \cdot \textbf{A} \cdot \textbf{I} + Q I_z^2 + g_n\mu_n \textbf{B} \cdot \textbf{I},
\end{equation}
where $D \approx 2.87$ GHz is zero-field splitting parameter, $\mu_e$ and $\mu_n$ are corresponding magnetons, and $g_e$ and $g_n$ are the $g$-factors, respectively. The ground state is splitted by $D$ due to spin-spin interaction between the unpaired electrons. Additional structure due to coupling to $^{14}$N nucleus is shown with $m_I = 0$ level separated from $m_I = \pm 1$, mainly caused by nuclear quadrupole interaction $Q$ and hyperfine splitting tensor $\textbf{A}$. $g_e \mu_e \textbf{B} \cdot \textbf{S}$ represents the electron Zeeman effect. In experiment, the static magnetic field along the NV axis $B_z$ is set to be 52 mT, so optical polarization of the nuclear spin can be easily realized.

The overall idea of our experiment is: to prepare a superposition state, find a suitable unitary $\mathcal{U}$ (represented by two MW pulses with the parameters of $B_1, B_2, \phi_1$ and $\phi_2$) using quantum autoencoder based on HQCA and the gradient descent algorithm, and realize the compression of two-qubit entanglement onto the $^{14}$N nuclear spin. The unitary for applying MW$_1$ and MW$_2$ consequently is given by
\renewcommand{\theequation}{S2}
\begin{equation}
\mathcal{U} = e^{-i2\pi t B_2 (\text{cos} \phi_2 \frac{\sigma_x}{2} \otimes \left | 0 \right \rangle \left \langle 0 \right | + \text{sin} \phi_2 \frac{\sigma_y}{2} \otimes \left | 0 \right \rangle \left \langle 0 \right |)} e^{-i2\pi t B_1 (\text{cos} \phi_1 \frac{\sigma_x}{2} \otimes \left | 1 \right \rangle \left \langle 1 \right | + \text{sin} \phi_1 \frac{\sigma_y}{2} \otimes \left | 1 \right \rangle \left \langle 1 \right |)}.
\end{equation}

In experiment, we choose $m_S=0$ and $m_S=-1$ in ground spin-triplet states as $\left | 0 \right \rangle$ and $\left | 1 \right \rangle$ of the electron spin qubit, and $m_I=0$ and $m_I=+1$ as $\left | 0 \right \rangle$ and $\left | 1 \right \rangle$ of the $^{14}$N nuclear spin qubit. With the pumping of 532 nm laser, the NV center can be polarized to $\left | 01 \right \rangle$ state. Hyperfine splitting (-2.153228 MHz) and the frequencies of RF$_1$ (5.101870 MHz) and RF$_2$ (2.940878 MHz) are calibrated via ODMR and electron-nuclear double resonance (ENDOR) experiment.

\section{two-qubit state tomography}
The partial density matrix $\rho$ of two-qubit entanglement is as follows, where we consider all four diagonal elements and off-diagonal elements $\left | 00 \right \rangle \left \langle 11 \right |$ and $\left | 11 \right \rangle \left \langle 00 \right |$.
\renewcommand{\theequation}{S3}
\begin{equation}
\rho = {
	\bordermatrix{%
		& \left \langle 00 \right | & \left \langle 01 \right | & \left \langle 10 \right | & \left \langle 11 \right | \cr
		\left | 00 \right \rangle & a_2 & \cdot & \cdot & \tilde{b} + i \tilde{c} \cr
		\left | 01 \right \rangle & \cdot & \tilde{a} & \cdot & \cdot \cr
		\left | 10 \right \rangle & \cdot & \cdot & a & \cdot \cr
		\left | 11 \right \rangle & \tilde{b} - i \tilde{c} & \cdot & \cdot & d
}}
\end{equation}
Here $a_2$, $\tilde{a}$, $a$ and $d$ represent the probability of $\left | 00 \right \rangle, \left | 01 \right \rangle, \left | 10 \right \rangle$ and $\left | 11 \right \rangle$ respectively, which relate directly to the fluorescence intensity. The first step of diagonal tomography is to calibrate the reference brightness of four pure two-qubit states, called $PL_{00}$, $PL_{01}$, $PL_{10}$ and $PL_{11}$. The experimental sequences for calibration is shown in Fig. \ref{figs2}(a).
\renewcommand{\thefigure}{S2}
\begin{figure}[ht]
	\centering
	\includegraphics[width=0.6\textwidth]{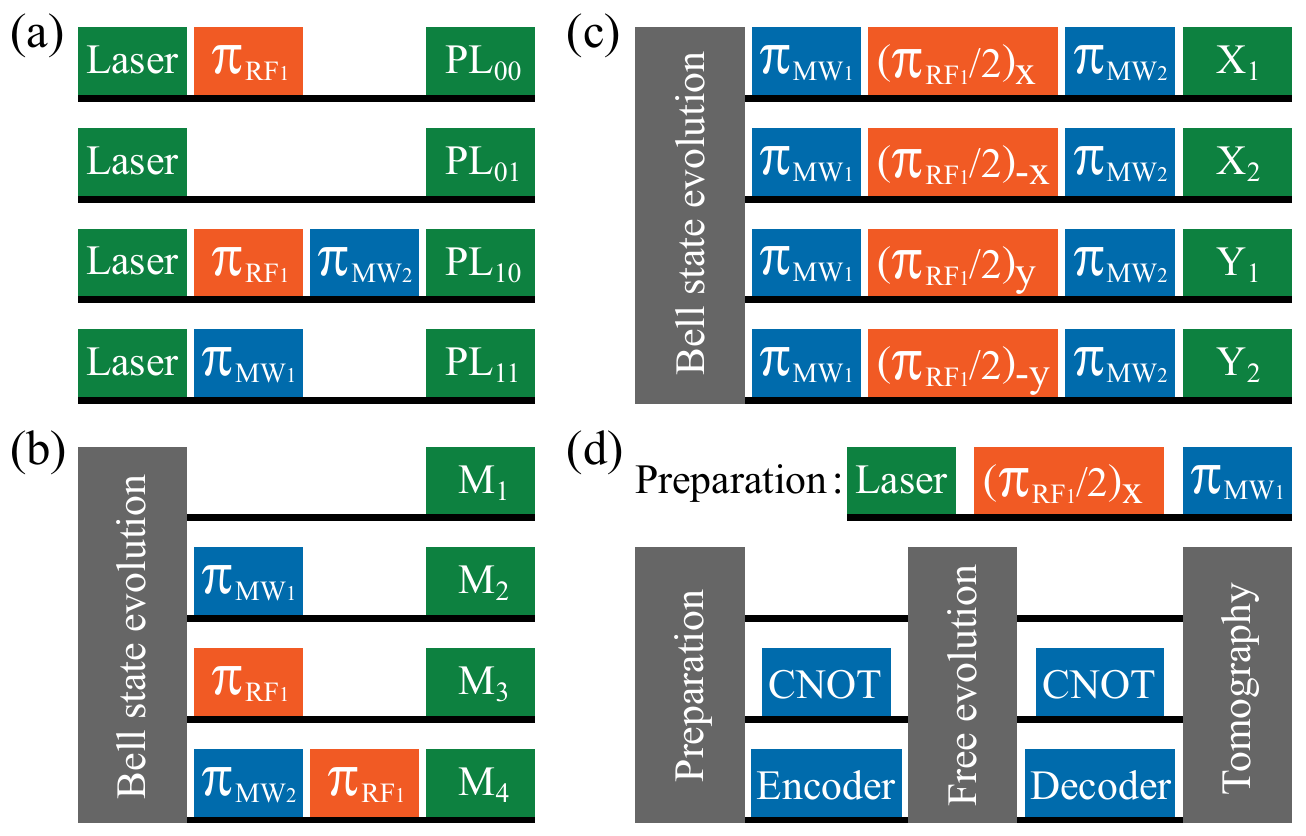}
	\caption{(a) Calibration of fluorescence intensity of four pure two-qubit states. (b-c) Tomography sequences of diagonal elements and off-diagonal element $\left | 00 \right \rangle \left \langle 11 \right |$, respectively. (d) Experimental sequences of preparing Bell state and three comparative experiments.}
	\label{figs2}
\end{figure}

For the diagonal tomography, there are four different experimental sequences, corresponding to the four measured values $M_1$, $M_2$, $M_3$ and $M_4$, as shown in Fig. \ref{figs2}(b). Then we calculate four diagonal elements according to the following equation
\renewcommand{\theequation}{S4}
\begin{equation}
\left( \begin{array}{cccc}
PL_{00}&PL_{01}&PL_{10}&PL_{11}\\
PL_{00}&PL_{11}&PL_{10}&PL_{01}\\
PL_{01}&PL_{00}&PL_{10}&PL_{11}\\
PL_{00}&PL_{10}&PL_{01}&PL_{11}
\end{array} \right)
\left( \begin{array}{c}
a_2\\
\tilde{a}\\
a\\
d
\end{array} \right)
= \left( \begin{array}{c}
M_1\\
M_2\\
M_3\\
M_4
\end{array} \right).
\end{equation}

As for measuring the lifetime of coherence $\left | 00 \right \rangle \left \langle 11 \right |$ or $\left | 11 \right \rangle \left \langle 00 \right |$ in the Bell state $\frac{1}{\sqrt{2}}(\left | 00 \right \rangle + \left | 11 \right \rangle)$, the off-diagonal tomography is necessary. The experimental sequences is shown in Fig. \ref{figs2}(c). Firstly we apply a MW$_1$ $\pi$ pulse, and the partial density matrix evolves as follows:
\renewcommand{\theequation}{S5}
\begin{equation}
\rho ={
	\left( \begin{array}{cccc}
	1&\cdot&\cdot&\cdot\\
	\cdot&0&\cdot&i\\
	\cdot&\cdot&1&\cdot\\
	\cdot&i&\cdot&0
	\end{array} \right)
	\left( \begin{array}{cccc}
	a_2&\cdot&\cdot&\tilde{b}+i\tilde{c}\\
	\cdot&\tilde{a}&\cdot&\cdot\\
	\cdot&\cdot&a&\cdot\\
	\tilde{b}-i\tilde{c}&\cdot&\cdot&d
	\end{array} \right)
	\left( \begin{array}{cccc}
	1&\cdot&\cdot&\cdot\\
	\cdot&0&\cdot&-i\\
	\cdot&\cdot&1&\cdot\\
	\cdot&-i&\cdot&0
	\end{array} \right)
} ={
	\left( \begin{array}{cccc}
	a_2&\tilde{c}-i\tilde{b}&\cdot&\cdot\\
	\tilde{c}+i\tilde{b}&d&\cdot&\cdot\\
	\cdot&\cdot&a&\cdot\\
	\cdot&\cdot&\cdot&\tilde{a}
	\end{array} \right)
}.
\end{equation}

Applying four RF$_1$ $\frac{\pi}{2}$ pulses with different phases, the evolutions of $2 \times 2$ density matrices ${\rho^{sub}_{1, 2, 3, 4}}$ go as follows:
\renewcommand{\theequation}{S6}
\begin{equation}
\rho_{1}^{sub} ={ \frac{1}{2}
	\left( \begin{array}{cc}
	1&i\\
	i&1
	\end{array} \right)
	\left( \begin{array}{cc}
	a_2&\tilde{c}-i\tilde{b}\\
	\tilde{c}+i\tilde{b}&d
	\end{array} \right)
	\left( \begin{array}{cc}
	1&-i\\
	-i&1
	\end{array} \right)
} ={
	\left( \begin{array}{cc}
	\frac{(a_2+d)}{2}-\tilde{b}&\tilde{c}+i\frac{(-a_2+d)}{2}\\
	\tilde{c}+i\frac{(a_2-d)}{2}&\frac{(a_2+d)}{2}+\tilde{b}
	\end{array} \right)
},
\end{equation}
\renewcommand{\theequation}{S7}
\begin{equation}
\rho_{2}^{sub} ={ \frac{1}{2}
	\left( \begin{array}{cc}
	1&-i\\
	-i&1
	\end{array} \right)
	\left( \begin{array}{cc}
	a_2&\tilde{c}-i\tilde{b}\\
	\tilde{c}+i\tilde{b}&d
	\end{array} \right)
	\left( \begin{array}{cc}
	1&i\\
	i&1
	\end{array} \right)
} ={
	\left( \begin{array}{cc}
	\frac{(a_2+d)}{2}+\tilde{b}&\tilde{c}+i\frac{(a_2-d)}{2}\\
	\tilde{c}+i\frac{(-a_2+d)}{2}&\frac{(a_2+d)}{2}-\tilde{b}
	\end{array} \right)
},
\end{equation}
\renewcommand{\theequation}{S8}
\begin{equation}
\rho_{3}^{sub} ={ \frac{1}{2}
	\left( \begin{array}{cc}
	1&-1\\
	1&1
	\end{array} \right)
	\left( \begin{array}{cc}
	a_2&\tilde{c}-i\tilde{b}\\
	\tilde{c}+i\tilde{b}&d
	\end{array} \right)
	\left( \begin{array}{cc}
	1&1\\
	-1&1
	\end{array} \right)
} ={
	\left( \begin{array}{cc}
	\frac{(a_2+d)}{2}-\tilde{c}&\frac{(a_2-d)}{2}-i\tilde{b}\\
	\frac{(a_2-d)}{2}+i\tilde{b}&\frac{(a_2+d)}{2}+\tilde{c}
	\end{array} \right)
},
\end{equation}
\renewcommand{\theequation}{S9}
\begin{equation}
\rho_{4}^{sub} ={ \frac{1}{2}
	\left( \begin{array}{cc}
	1&1\\
	-1&1
	\end{array} \right)
	\left( \begin{array}{cc}
	a_2&\tilde{c}-i\tilde{b}\\
	\tilde{c}+i\tilde{b}&d
	\end{array} \right)
	\left( \begin{array}{cc}
	1&-1\\
	1&1
	\end{array} \right)
} ={
	\left( \begin{array}{cc}
	\frac{(a_2+d)}{2}+\tilde{c}&\frac{(-a_2+d)}{2}-i\tilde{b}\\
	\frac{(-a_2+d)}{2}+i\tilde{b}&\frac{(a_2+d)}{2}-\tilde{c}
	\end{array} \right)
}.
\end{equation}

At this time, the off-diagonal elements are transfered to the diagonal elements and its optical readout is allowed. Following a MW$_2$ $\pi$ pulse and optical readout, we have:
\renewcommand{\theequation}{S10}
\begin{equation}
X_1 = aPL_{00} + (\frac{a_2+d}{2}+\tilde{b})PL_{01} + (\frac{a_2+d}{2}-\tilde{b})PL_{10} + \tilde{a}PL_{11},
\end{equation}
\renewcommand{\theequation}{S11}
\begin{equation}
X_2 = aPL_{00} + (\frac{a_2+d}{2}-\tilde{b})PL_{01} + (\frac{a_2+d}{2}+\tilde{b})PL_{10} + \tilde{a}PL_{11},
\end{equation}
\renewcommand{\theequation}{S12}
\begin{equation}
Y_1 = aPL_{00} + (\frac{a_2+d}{2}+\tilde{c})PL_{01} + (\frac{a_2+d}{2}-\tilde{c})PL_{10} + \tilde{a}PL_{11},
\end{equation}
\renewcommand{\theequation}{S13}
\begin{equation}
Y_2 = aPL_{00} + (\frac{a_2+d}{2}-\tilde{c})PL_{01} + (\frac{a_2+d}{2}+\tilde{c})PL_{10} + \tilde{a}PL_{11},
\end{equation}
\renewcommand{\theequation}{S14}
\begin{equation}
\tilde{b} = \frac{X_1-X_2}{2(PL_{01}-PL_{10})},
\end{equation}
\renewcommand{\theequation}{S15}
\begin{equation}
\tilde{c} = \frac{Y_1-Y_2}{2(PL_{01}-PL_{10})}.
\end{equation}

In this work, we use the modulus of off-diagonal elements $\sqrt{\tilde{b}^2 + \tilde{c}^2}$ to measure the coherence of the Bell state.

\section{Multi-qubit entangled states}
It is known that universal quantum gates can be realized in terms of single-qubit rotations and two-qubit CNOT gates, hence, a straightforward idea is to decompose a desired gate into these two kinds of gates. To show the polynomial growth of parameters, we consider a more specific decomposition where the CNOT gates are limited to two-qubit controlled rotations, which simplifies the calculations without loss of generality. For an $ n $-qubit quantum circuit, applying $ n $ independent rotations on each qubit at the beginning and the end of the circuit yields $ (2\times 3n) $ parameters, since any single-qubit unitary transformation can be characterized by three Euler angles. Then, $ n(n-1) $ two-qubit controlled-rotation gates are inserted between single-qubit rotations, corresponding to that each qubit as the control qubit has $ (n-1) $ possible controlled gates with $ (n-1) $  qubits as target qubits, respectively. By this means, the number of parametrized gates is reduced to $ n_g = n(n-1)+ 2 n $ for $ n $ qubits, while the number of total parameters becomes $ 3n_g $.
\renewcommand{\thefigure}{S3}
\begin{figure}[ht]
	\centering
	\includegraphics{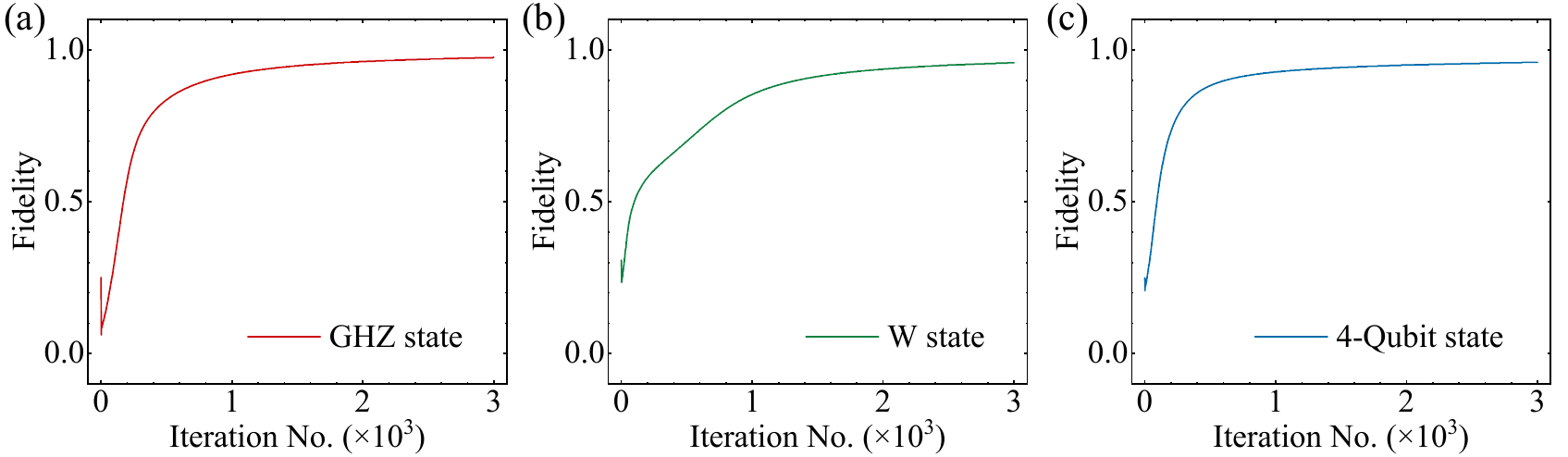}
	\caption{Simulation results of multi-qubit autoencoder. For three-qubit state (a) GHZ state: $ |\mathrm{GHZ}\rangle = \left(|000\rangle+|111\rangle\right)/\sqrt{2} $, (b) W state: $ |W\rangle = \left(|001\rangle+|010\rangle+|100\rangle\right)/\sqrt{3} $ and the four-qubit state (c), whose coefficients are $ \sqrt{i/136}\ (i= 1,2,3,\dots,16) $ for $ i $th basis of $ 16 $ basis states respectively, the fidelities converge rapidly in first thousand iterations.}
	\label{figs3}
\end{figure}

To demonstrate the availability of autoencoder for multi-qubit quantum circuit, in addition to the two-qubit experiment, we also simulated 3-qubit and 4-qubit cases, as shown in Fig. \ref{figs3}.
We chose the GHZ state $ |\mathrm{GHZ}\rangle = \left(|000\rangle+|111\rangle\right)/\sqrt{2} $ and W state $ |W\rangle = \left(|001\rangle+|010\rangle+|100\rangle\right)/\sqrt{3} $, which are both of great physical interests in quantum information science, to be the input state of a 3-qubit circuit, respectively. The fidelity of compressing GHZ (W) state into 1 qubit is up to $ 0.9749 \ (0.9578) $, respectively. In the case of 4-qubit, we chose a more general state whose coefficients of the eigenstates $\{|0000\rangle, |0001\rangle, \cdots, |1111\rangle\}$ are respectively $\{\sqrt{1/136}, \sqrt{2/136}, \cdots, \sqrt{16/136}\}$, and compressed the state to 2 qubits with fidelity 0.9592.

\end{document}